\documentclass[aps,prl,twocolumn,showpacs,preprintnumbers]{revtex4}
\usepackage{blindtext}
\usepackage{xcolor}

\usepackage{amssymb,amsmath}
\usepackage{graphicx}
\usepackage{upgreek}
\usepackage[english]{babel}
\usepackage[T1]{fontenc}
\usepackage[utf8]{inputenc}
\usepackage{graphicx}
\usepackage{bm}

\DeclareMathOperator{\sgn}{sgn}

\begin{document}

\title{Observation of topological gravity-capillary waves in a water wave crystal}

\author{Nicolas Laforge}
\author{Vincent Laude}
\author{Franck Chollet}
\author{Abdelkrim Khelif}
\author{Muamer Kadic}
\email{muamer.kadic@femto-st.fr}
\address{Institut FEMTO-ST, CNRS UMR 6174, Universit\'{e} de Bourgogne Franche-Comt\'{e}, Besan\c{c}on, France}

\author{Yuning Guo}
\author{Romain Fleury}
\address{Laboratory of Wave Engineering (LWE), Ecole Polytechnique Fédérale de Lausanne (EPFL), Switzerland}


\begin{abstract}
The discovery of topological phases of matter, initially driven by theoretical advances in quantum condensed matter physics, has been recently extended to classical wave systems, reaching out to a wealth of novel potential applications in signal manipulation and energy concentration. Despite the fact that many realistic wave media (metals at optical frequencies, polymers at ultrasonic frequencies) are inherently dispersive, topological wave transport in photonic and phononic crystals has so far been limited to ideal situations and proof-of-concept experiments involving dispersionless media. 
Here, we report the first experimental demonstration of topological edge states in a classical water wave system supporting highly dispersive wave propagation, in the intermediate regime of gravity-capillary waves. We use a stochastic method to rigorously take into account the inherent dispersion and devise a water wave crystal insulator supporting valley-selective transport at topological domain walls. Our measurements, performed with a high-speed camera under stroboscopic illumination, unambiguously demonstrate the possibility of valley-locked transport of water waves.
\end{abstract}


\maketitle

Topological insulators are bulk insulators whose bands are characterized by a quantized number known as a topological invariant \cite{Hafezi2014,Silveirinha2015}, which cannot change upon continuous transformations of the band structure.
This topological property of the bands implies the presence of edge states at topological interfaces, which is protected by the topology of the surrounding bulk insulators \cite{Zhu2018,Craster2018,Chen2014b,Ma2015,Vila2017,Lu2017,Shen2019,He2016}.
Originally discovered in condensed matter systems, including Quantum-Hall \cite{Haldane2008,Wang2008,Hafezi2011,Klitzing1980,Thouless1982,Haldane1988,Hatsugai1993,Ohgushi2000,Ren2016} and Quantum Spin-Hall insulators \cite{Kane2005a,Kane2005b,Bernevig2006a,Bernevig2006b}, the concept of topological transport has recently been transposed to various fields of classical wave physics, including optics \cite{Khanikaev2012,Hafezi2013,Lu2014}, acoustics \cite{Lu2017,Peng2016,Khanikaev2015,Yang2015}, microwaves \cite{Wang2008,Gao2016}, and mechanics \cite{Nash2015,Huber2015,Huber2016,Huber2017,Shen2019}, where it represents a promising way to transport signals and concentrate energy in a robust, symmetry-protected way.
While classical analogs of Chern insulators \cite{Raghu2008,Khanikaev2015}, quantum spin-Hall systems \cite{Hsieh2008}, and valley-Hall insulators \cite{Lu2016,Lu2017,Prodan2018b} have been previously studied and demonstrated, prior arts have focused mainly on idealistic situations in which the dispersion of the host materials have been neglected or avoided.
This drastic assumption, however, holds only for a small subset of the available physical platforms in which exploiting topological physics could have large practical implications. 
It does not hold, for instance, for water wave systems, which generally support highly dispersive surface waves \cite{Miles1978}.
Yet, controlling the energy carried by ocean waves, and forcing it to concentrate at a location where it can be harvested, would be a fascinating application of topological physics, providing topological edge modes are compatible with the highly dispersive character of these systems. 

\begin{figure}[!tbh]
\includegraphics[width=8.6cm]{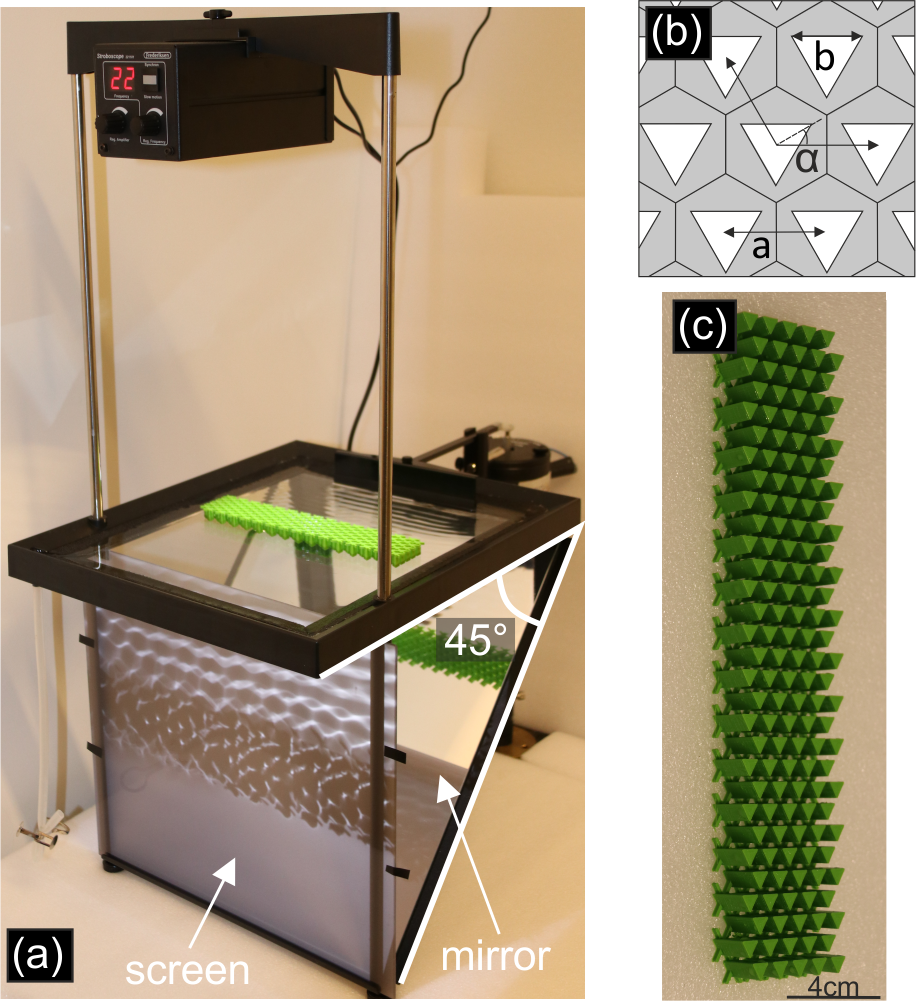}
\caption{Experimental setup for observation of topological gravity-capillary waves in a water wave crystal.
(a) The crystal sample (green color) is placed in a water tank.
A mechanical straight paddle is exciting vertical motion of the water surface at the same frequency as the stroboscopic illumination.
A mirror is placed at 45$^\circ$ below and reflects the image on a diffusive screen placed in front.
(b) The hexagonal crystal is made of triangular pillars with $a=8$~mm, $b=6.4$~mm, and a variable orientation angle $\alpha$ (here $\alpha=30^\circ$).
(c) A typical 3D-printed crystal sample made of PLA (polylactic acid) is shown.}
\label{Figure1}
\end{figure}

In this Letter, we demonstrate experimentally the relevance of topological physics in a classical wave system with strong dispersion, namely gravity-capillary waves at a water-air interface interacting with a water wave crystal.
We use a stochastic method to obtain the dispersion relation in the crystal, using Bloch's theorem, and we design topological edge states based on valley conservation.
Our measurements, based on direct imaging using a high-speed camera under stroboscopic illumination (see Fig.~\ref{Figure1}), demonstrate unambiguously the possibility of topological transport in systems with very strong dispersion, extending the reach of topological physics to a wealth of new physical platforms.

\bigskip
Gravity-capillary water waves are surface waves resulting from the balance of the potential energy of gravity forces and surface tension with the kinetic energy of a water column.
As a result of body (volume) and interface (surface) contributions, their velocity is isotropic but inherently strongly dispersive and dependent on the water depth.
In order to describe the propagation of gravity-capillary waves in the regime of small water elevation (as compared to the water depth), we consider the linear velocity potential theory for water waves~\cite{Hu2003,Hu2004,Jeong2004}.
In the absence of external forces the vertical displacement (elevation) of the liquid-air interface $\eta(x,y)$ satisfies the two-dimensional partial differential equation
\begin{equation}
\label{MSE}
\nabla \cdot (c_\mathrm{p} c_\mathrm{g}\nabla \eta) + \kappa^2 c_\mathrm{p} c_\mathrm{g} \eta = 0,
\end{equation}
where $c_\mathrm{p}$ is the phase velocity, $c_\mathrm{g}$ is the group velocity, and $\kappa$ is the wavenumber.
A similar equation is satisfied by the horizontal part of the velocity potential~\cite{Hu2003}.
We further recall that the dispersion relation between wavenumber and angular frequency $\omega$ for a horizontal liquid-air interface is given by
\begin{equation}
\label{DR}
\omega^2 = g \kappa (1 + d_c^2 \kappa^2) \tanh(\kappa h),
\end{equation}
where $g$ is the gravitational acceleration, $h$ is the water depth, and $d_c$ is the capillary length.
The phase velocity is then 
\begin{equation}
\label{CP}
c_\mathrm{p} = \frac{\omega}{\kappa} = \sqrt{\frac{g}{\kappa}(1 + d_c^2 \kappa^2)\tanh(\kappa h)}
\end{equation}
and the group velocity is
\begin{equation}
\label{CG}
c_\mathrm{g} = \frac{d\omega}{d\kappa} = \left( \frac{1}{2} + \frac{d_c^2 \kappa^2}{1 + d_c^2 \kappa^2} + \frac{\kappa h}{\sinh(2\kappa h)} \right) c_\mathrm{p} ,
\end{equation}
where the capillary length is given by 
$d_c = \sqrt{\frac{\gamma}{g\rho}}$, with $\gamma$ the surface tension and $\rho$ the mass density of the liquid.

\begin{figure}[!t]
\includegraphics[width=8.6cm]{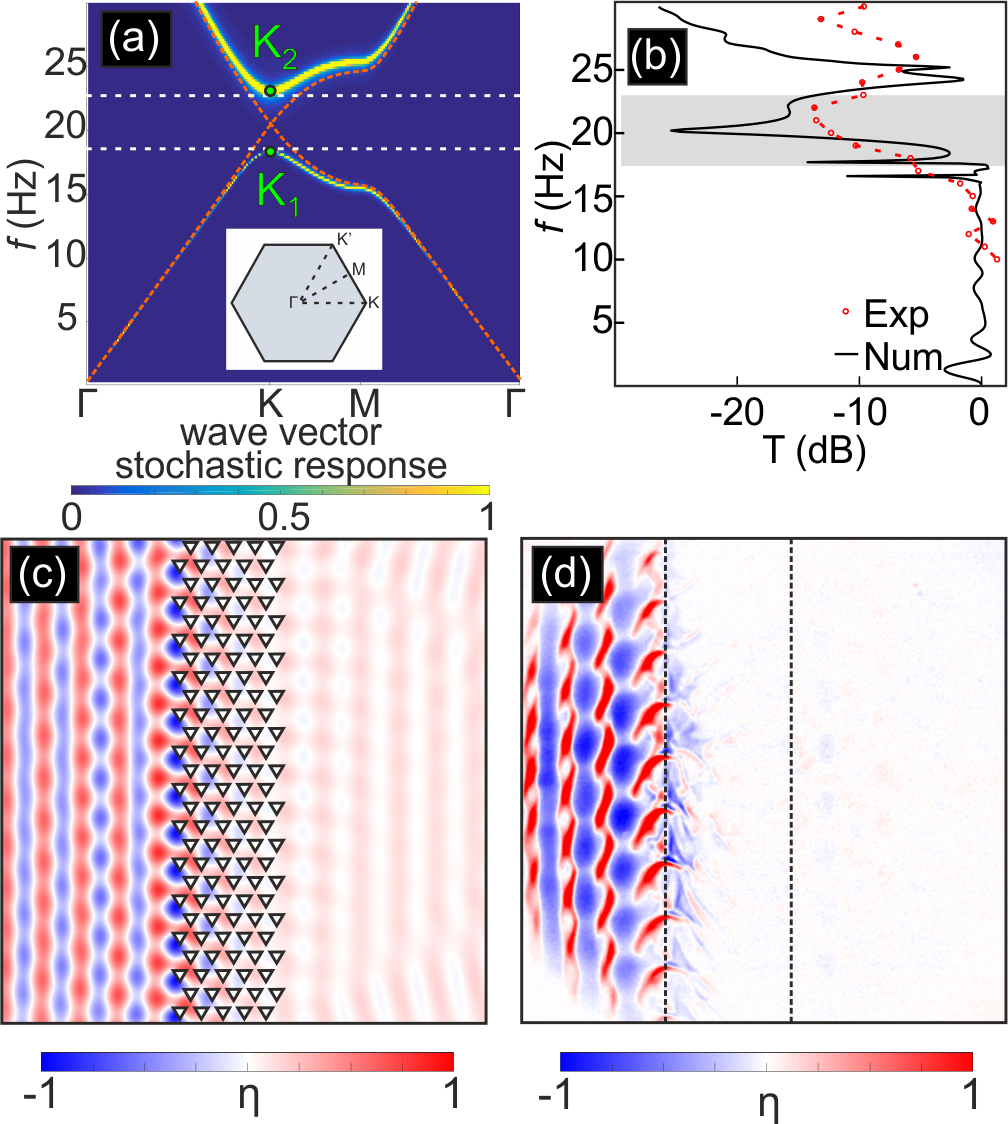}
\caption{Phononic properties of the water wave crystal of Fig.~\ref{Figure1}.
(a) The dispersion relation is obtained as the response to a stochastic excitation~\cite{Laude2018} for a water level $h = 7.3$~mm and $\alpha=30^o$ (the dispersion relation for $\alpha=0^o$ is the red line).
The inset represents the first Brillouin zone (BZ).
(b) The measured transmission of a plane wave through a 5 unit cell thick crystal slab (dotted line) is compared with the numerically computed transmission (solid line).
(c) Numerical simulation of the reflection of a plane wave at $22$ Hz (inside the phononic band gap) on the crystal. 
(d) Experimentally recorded surface elevation at the same frequency.
The boundaries of the crystal are indicated with dashed lines.
}
\label{Figure2}
\end{figure}

Inside the artificial crystal sample, Eq.~\eqref{MSE} can be used to obtain Bloch waves and therefore band structures.
Indeed, Eq.~(\ref{MSE}) can be recast as a Helmholtz equation for scalar waves in a dispersive medium
\begin{equation}
\label{MSE2}
\nabla \cdot (A(\omega) \nabla \eta) + \omega^2 B(\omega) \eta = 0,
\end{equation}
with coefficients $A(\omega)=c_\mathrm{p} c_\mathrm{g}$ and $B(\omega)=c_\mathrm{g} / c_\mathrm{p}$ depending explicitly on frequency.
A Neumann boundary condition is considered at the interfaces between pillars and water, i.e. $\partial \eta / \partial n=0$, where $n$ is the normal to the interface~\cite{Hu2003,Hu2004,Jeong2004}.
Bloch waves have the form $\eta(x,y) = \eta_{\mathbf{k}}(x,y) \exp(\imath \mathbf{k} \cdot \mathbf{r})$, with $\mathbf{k}$ the Bloch wavevector and $\eta_{\mathbf{k}}$ the periodic part of the Bloch wave.
Computing eigenvalues and eigenfunctions to obtain the Bloch waves of the crystal is not straightforward since the coefficients of the equation depend on frequency.
We use the stochastic excitation method proposed by Laude and Korotyaeva~\cite{Laude2018}.
The method considers all possible values of frequency $\omega$ and Bloch wavevector $\mathbf{k}$ and observes the response of the system to a spatially random force distributed inside the unit-cell.
The finite element implementation for Eq.~\eqref{MSE2} follows the prescriptions in Ref.~\cite{Laude2018}.
A further difficulty is that the model does not take into account certain interface effects between water and the solid pillars, including the formation of a meniscus as a result of capillary forces.
As the distance between pillars is smaller than the wavelength \cite{Dupont2015}, such effects can be approximated by using an effective dispersion relation~\cite{supp}.
In practice, the effective value of the surface tension is adjusted according to experimental observations.
The derived value, $\gamma_{\rm eff} = 0.17$~N$\cdot$m$^{-1}$, is larger than the usual value for the unperturbed water-air interface, corresponding to an effective increase of the phase velocity inside the artificial crystal.
The phenomenological model for the artificial crystal is thus composed of Eq.~\eqref{MSE2} together with the effective dispersion relation in Eq.~\eqref{DR} and Neumann boundary conditions at the pillars.

We compute phononic band structures $\omega(\mathbf{k})$ along the boundary of the irreducible Brillouin zone (BZ) (namely the path $\rm \Gamma- K- M- \Gamma$) as depicted in Fig.~\ref{Figure2}(a) for the geometry of Fig.~\ref{Figure1}(b).
Phononic band structures are shown for an hexagonal crystal of triangular pillars with lattice constant $a = 8$~mm and $b/a = 0.8$. 
The phononic band gap extends from 18 to 23~Hz for $\alpha=30^\circ$, whereas the band gap closes for $\alpha=0^\circ$.
The transmission through a finite crystal sample composed of 5 periods ($\alpha=30^\circ$) is shown as a function of frequency in Fig.~\ref{Figure2}(b) for direction $\Gamma$K.
A good agreement between theory and experiment is observed.
At a frequency of 22~Hz, inside the phononic band gap, incident waves are indeed reflected by our crystalline insulator and form a standing wave pattern on the incident side, as illustrated in Fig.~\ref{Figure2}(c).
Full wave numerics are obtained by solving Eq.~(\ref{MSE2}) using the finite element method in Comsol Multiphysics at a fixed frequency.
Fig~\ref{Figure2}(c) shows an example of the water wave field obtained using this method.
PMLs (perfectly matched layers) are used to absorb the outgoing waves.

\begin{figure}[!tb]
\includegraphics[width=8.6cm,angle=0]{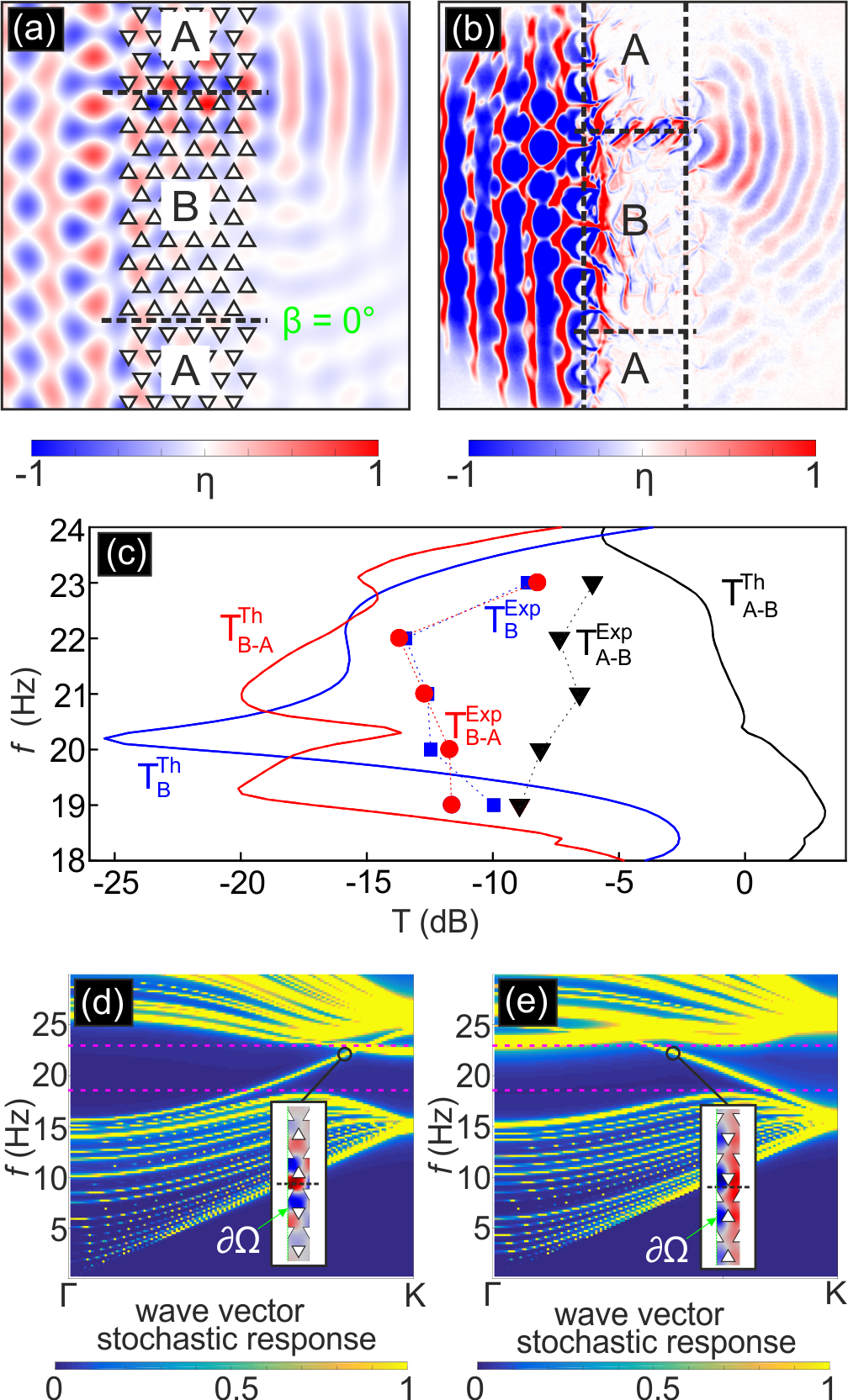}
\caption{Valley-selective excitation of water wave edge modes for two interfaces at $\beta=0^\circ$. (a) Numerical simulation and (b) experimental measure of transmission at 22 Hz. (c) Quantitative comparison of the transmission for A-B and B-A interfaces in the experimental and numerical cases, as well as the case of the domain B. Dots are used for experimentally measured points. Dispersion relations obtained for supercells for the interface B-A in (d) and A-B in (e).}
\label{Figure3}
\end{figure}

The experimental setup in Fig.~\ref{Figure1} is composed of a $24\times30$ cm$^2$ water wave ripple tank illuminated by stroboscopic light synchronized with a straight paddle, operating at a frequency tunable between 10 and 75 Hz.
Shadows caused by light refraction at the sinusoidally modulated water surface are formed onto a screen after reflection off a mirror and are recorded with a camera.
A typical experimental wave pattern is shown in Fig.~\ref{Figure1}(a), without any post-processing.
The observed patterns are related to the local surface curvature (a bright fringe indicating a positive curvature and a local elevation of the water level).
Hence, after image processing, we obtain a quantity proportional to the vertical elevation of the water surface.
Crystal samples are fabricated with 3D additive printing in PLA and a typical sample is shown in Fig.~\ref{Figure1}(b).
Samples are composed of pillars that are higher than the water level (15 mm, and the water level without the crystal is equal to 7 mm).
The experimental transmission for $\alpha$ = $30^\circ$ is shown as a function of frequency in Fig.~\ref{Figure2}(b) and is in fair agreement with theory.
The experimental image in Fig.~\ref{Figure2}(d) confirms the strong phononic band gap reflection near 22 Hz.

\begin{figure}[!tb]
	\includegraphics[width=8.6cm,angle=0]{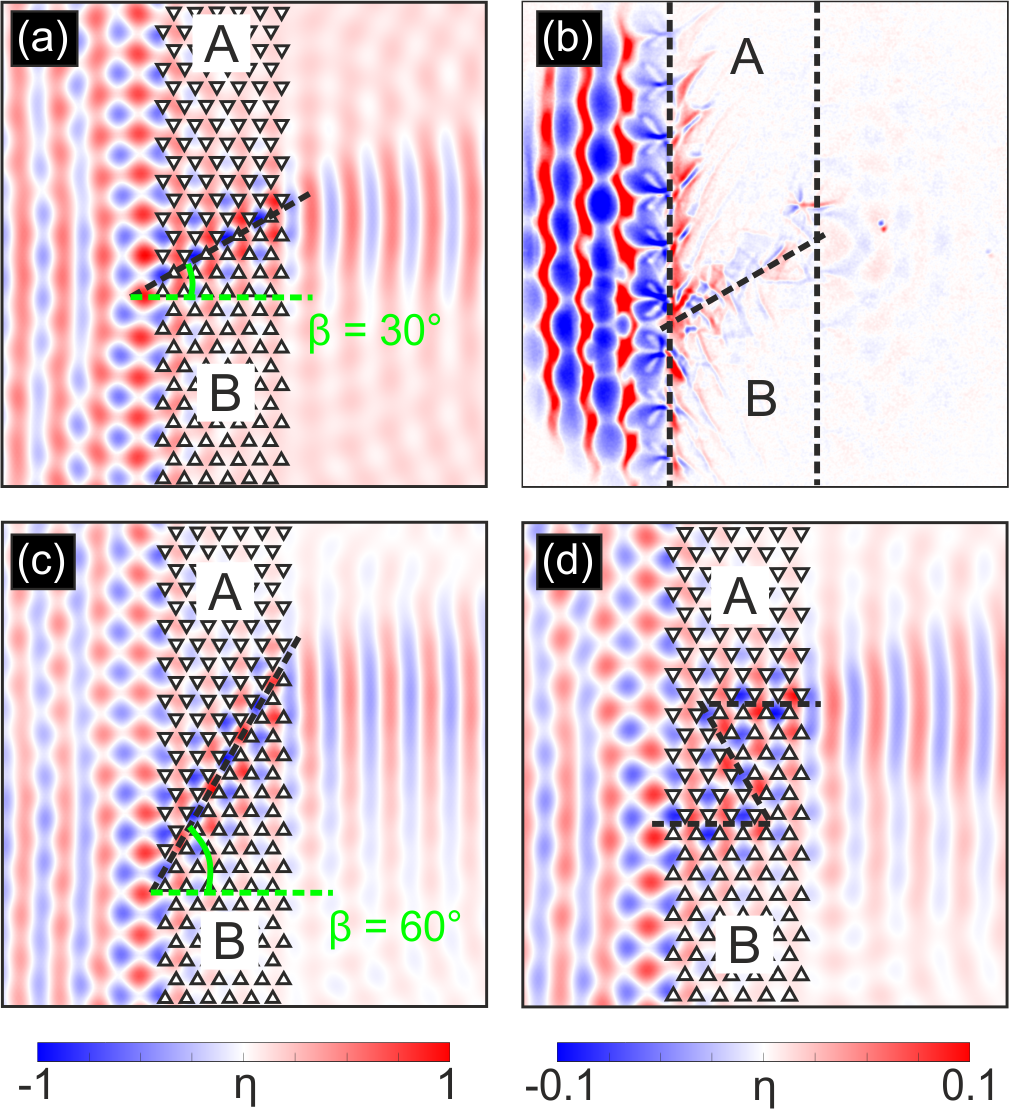}
	\caption{Water wave fields obtained for different geometries of interfaces at 22~Hz. (a) Numerical simulation and (b) experimental transmission of a $30^\circ$ interface. (c) Numerical simulation of a $60^\circ$ interface and (d) Numerical simulation of a zigzag shaped interface.}
	\label{Figure4}
\end{figure}

The 2D hexagonal crystal of rotated opaque triangles supports valley vortex Bloch waves carrying a quantized topological phase~\cite{Lu2016}.
Actually, when $\alpha=0^\circ$, the plane crystallographic group -- or \textit{wallpaper group} -- of the crystal is \textit{\textbf{p31m}}, meaning that there are three 3-fold axes of rotation and three reflection planes, two of which are images in a reflection.
When $\alpha \neq 0$, the symmetry is reduced to that of wallpaper group \textit{\textbf{p3}}, with only three 3-fold axes of rotation remaining~\footnote{For $\alpha=30^\circ$, the wallpaper group is \textit{\textbf{p3m1}}, and there are three 3-fold axes of rotation and three reflection planes intersecting at the rotation centers.}.
The Bloch waves defining the phononic band gap follow the same \textit{\textbf{p3}} symmetry and show three vortices placed either in between the vertices of the triangles (Bloch waves $\rm K_1$, with valley topological phase $+\sgn(\alpha)$) or in between the sides of the triangles (Bloch waves $\rm K_2$, with valley topological phase $-\sgn(\alpha)$)~\cite{Lu2016}; see Fig.~S2 of the Supplemental Material for a representation of the $\rm K_1$ and $\rm K_2$ Bloch waves~\cite{supp}.
Note that the vortex centers actually coincide with rotation centers of the crystal.

The crystal B of triangles rotated by $-|\alpha|$ is the image in a \textit{glide reflection} of axis $Ox$ and translation $a/2$ of the crystal A of triangles rotated by $+|\alpha|$.
This property can be exploited to construct a domain wall (DW) separating the two non equivalent chiral crystals A and B.
As depicted in Fig.~\ref{Figure3}(a), the DW A-B with A placed above B is different from the DW B-A with B placed above A.
Both DWs have the symmetry of the \textit{frieze group} \textit{\textbf{p11g}}.
It was shown for classical non dispersive waves that the DWs support unidirectional edge waves~\cite{Lu2017}; here we show that the property remains true for dispersive water waves.
Both the numerical result in Fig.~\ref{Figure3}(a) and the corresponding experiment in Fig.~\ref{Figure3}(b) show that a plane surface wave incident normally from the left is funneled to the right side for the A-B DW, but not for the B-A DW.
The reason for this asymmetry is discussed below.
The transmission through the topological waveguides was estimated and is presented in Fig.~\ref{Figure3}(c); again experiment and numerics agree fairly well.

The edge modes are obtained numerically using the stochastic excitation method, considering a super-cell encompassing 5 rows of each crystals A and B.
The phononic band structures for DWs A-B and B-A are shown in Fig.~\ref{Figure3}(d-e).
In each case a edge mode appears inside the phononic band gap, traversing it to connect the systems of bulk bands extending below and above.
The A-B edge wave has a negative dispersion -- its group velocity is negative for positive wavenumbers, -- whereas the B-A edge wave has a positive dispersion.
The insets in the figures show the modal distribution of both edge waves.
It can actually be verified that the A-B edge wave is composed of $\rm K_1$ valley vortex waves that are evanescent in the transverse direction.
Conversely, the B-A edge wave is composed of $\rm K_2$ evanescent valley vortex waves.
As further discussed in the supplemental Material~\cite{supp}, the sign of the topological charge is in direct connection with the sign of the group velocity of the edge waves.

The possibility of coupling an externally incident plane wave with an edge wave of the domain wall can be evaluated by comparing their modal fields.
We evaluate the following overlap integral
\begin{equation}
\Psi = \frac{|\int_{\partial\Omega} \eta_0 \cdot \eta(x,y) \; \mathrm{d}y|}{\int_{\partial\Omega}|\eta(x,y)| \; \mathrm{d}y} .
\end{equation} computed on the interface $\partial \Omega$ between the crystal and the incidence region. 
The resulting number, which varies between 0 and 1, measures the matching of the edge mode with a plane wave with amplitude $\eta_0=1$ along the interface $\partial\Omega$.
We found numerically that $\Psi_{\rm A-B} \approx 1$ for the A-B DW, and that $\Psi_{\rm B-A} \approx 0.03$  for the B-A DW.
These numbers confirm the asymmetry of coupling of the normally incident plane wave to both edge modes.
Furthermore, the overlap integral computed for all Bloch modes~\cite{supp} confirms that transmission occurs only along DW A-B for all frequencies. 

A DW waveguide making an angle of $30^\circ$ with respect to the direction of propagation was also fabricated and tested (see Fig.~\ref{Figure4}(a-b)).
Again, theory and experiment agree and show that at the output interface a point-like source emission is obtained.
Note that the $30^\circ$-rotated DW is not purely of the A-B or B-A type and actually combines $\rm K_1$ and $\rm K_2$ vortex states to funnel surface waves.
The $60^\circ$-rotated DW shown in Fig.~\ref{Figure4}(c) is of the B-A type.
Due to non normal incidence the $\rm K_2$ vortex edge wave can be excited in this case.
Finally, the Z shape DW of Fig.~\ref{Figure4}(d) combines a series of $120^\circ$ turns and is of the A-B type.
However, the latter crystal has a very low confinement and was thus only tested numerically.

As a conclusion, we performed an experimental demonstration of the existence of topological edge states guided a domain wall of a water wave crystal, in the intermediate regime of gravity-capillary waves.
The edge states are a superposition of vortex waves carrying a quantized topological phase~\cite{Lu2016} and can be described by a classical analogy to the valley Hall effect~\cite{Lu2017,Pal2017,Vila2017}.
A good qualitative agreement between theory and experiments was obtained, with the capillary effects at the interface between water and the crystal modeled via the use of an effective parameter.
Our observations extend the reach of topological wave physics to a wide range of physical platforms containing highly dispersive media, not only water wave systems, but also elastic waves and plasmonics.

\section*{Acknowledgments}

We acknowledge the assistance of C.-L. Azzopardi, V. Pêcheur, J-F. Manceau, E. Carry, and T. Daugey for samples fabrication, F. Cherioux and S. Benchabane for discussions and A. Mosset for the measurement setup.
This work has been supported by the EIPHI Graduate School (contract ANR-17-EURE-0002) and by the French Investissements d'Avenir program, project ISITE-BFC (contract ANR-15-IDEX-03).



\end{document}